\documentclass[a4paper,12pt]{article}        
              
\usepackage{amsthm,amsfonts,amsmath,amssymb} 
\usepackage{mathrsfs}                        
\usepackage{dsfont}  
\usepackage{empheq}  
\usepackage{hyperref}        
\usepackage[T1]{fontenc}          
\usepackage[latin1]{inputenc}     

\usepackage[margin=7em]{geometry}    

\newcommand{\ud}{\mathrm{d}}

\newcommand{\e}{\mathrm{e}}
\newcommand{\CR}{\mathds{R}}
\newcommand{\CZ}{\mathds{Z}}
\newcommand{\CC}{\mathds{C}}

\newcommand{\dep}[2]{\displaystyle\frac{\partial\emph{$#1$}}{\partial{
\emph{$#2$}}}}
\newcommand{\depp}[2]{\tfrac{\partial\emph{$#1$}}{\partial{
\emph{$#2$}}}}

\newcommand{\fraction}{\displaystyle\frac}

\newcommand{\integrall}[3]{\int_{\emph{$#1$}}^{\emph{$#2$}}{\emph{$#3$}}}
\newcommand{\Somatorio}[2]{\displaystyle\sum\limits_{\emph{$#1$}}^
{\emph{$#2$}}}

\newcommand{\escalar}[2]{\left\langle{\emph{$#1$}},{\emph{$#2$}}\right\rangle}
\theoremstyle{plain} 
\newtheorem{teo}{Theorem}[section]
\newtheorem{defi}{Definition}[section]
\newtheorem{lem}{Lemma}[section]
\newtheorem{prop}{Proposition}[section]

\theoremstyle{definition}

\newtheorem{rema}{Remark}[section]

\begin{document}

\title{Charge quantisation without magnetic poles: a topological approach to electromagnetism}
\author{Romero Solha\footnote{Email address: romerosolha@gmail.com }}

\date{\today}

\maketitle


\begin{abstract}The present work provides a theoretical explanation for the quantisation of electric charges, an open problem since Millikan's oil drop experiment in 1909. This explanation is based solely on Maxwell's theory, it recasts Electromagnetic Theory under the language of complex line bundles; therefore, neither magnetic poles nor quantum mechanics are invoked. 
\end{abstract} 


\section{Introduction}

\hspace{1.5em}The existence of magnetic poles was essentially the only theoretical explanation for charge quantisation (e.g. Dirac's magnetic pole \cite{PaulDirac31}), and there is no experimental data supporting their existence.

My intention with this note is to clarify some features of the Electromagnetic Theory and to champion the idea that Maxwell's theory is not only intrinsically relativistic, but it naturally accommodates phenomena that are usually associated with its quantum aspects: charge quantisation, the Aharonov--Bohm effect, Dirac's equation, and photon wave functions. All of these become clear by defining the electromagnetic field (over a spacetime region) as an equivalence class of hermitian line bundles with hermitian connexion (considering the spacetime region of interest as the base manifold).

I shall justify my approach by quoting Dirac \cite{PaulDirac31}: ``The most powerful method of advance that can be suggested at present ...'', still is, I think, ``... to employ all the resources of pure mathematics in attempts to perfect and generalise the mathematical formalism that forms the existing basis of theoretical physics, and after each success in this direction, to try to interpret the new mathematical features in terms of physical entities''. 

An appendix to this note is included. It supplements the text with definitions and theorems (together with sketch of their proofs) that are known in the literature, providing a place to fix notation and a guide for the reader not familiar with differential topology. 

Throughout this note and otherwise stated, all the objects considered will be $C^\infty$; manifolds are real, Hausdorff, paracompact, connected, and orientable; and physical quantities are measured using the International System of Units. 

\subsection{Acknowledgements}

\hspace{1.5em}I want to express my gratitude to Ricardo S. Schor, Jonathan Weitsman, J\c{e}drzej \'{S}niatycki, and Yuri Ximenes Martins, for commenting on drafts of this note; as well as to Pierre-Louis de Assis, Rafael Rabelo, Emilson R. Viana, and Diogo D. dos Reis, for helpful discussions. This work was partially supported by PNPD/CAPES.


\section{A pr\'{e}cis of the Electromagnetic Theory}\label{PEMT}

\hspace{1.5em}Classical electromagnetism over a spacetime $(M,\boldsymbol{\mathrm{g}})$, a lorentzian manifold satisfying Einstein's field equations \cite{Wal}, is described by a differential $2$-form, the electromagnetic $2$-form $\omega\in\Omega^2(M;\CR)$, solving Maxwell's equations, 
\begin{equation*}
\left\{\begin{array}{c}\ud\omega=\star_{\boldsymbol{\mathrm{g}}}\alpha \\ \star_{\boldsymbol{\mathrm{g}}}\ud\star_{\boldsymbol{\mathrm{g}}}\omega=0 \end{array}\right. \ ,
\end{equation*}where $\star_{\boldsymbol{\mathrm{g}}}$ is the Hodge star operator of the lorentzian metric $\boldsymbol{\mathrm{g}}$, and the differential \newline $1$-form $\alpha\in\Omega^1(M;\CR)$ encodes the information about charge and current distributions. On the Minkowski spacetime, these equations are equivalent to\footnote{My choices are such that the electromagnetic $2$-form has dimension of charge (ampere times second in SI units).}: 
\begin{equation*}
\left\{\begin{array}{l}\vec{\nabla}\cdot\epsilon_0\vec{E}=\rho \\ \vec{\nabla}\times\tfrac{1}{\mu_0 c}\vec{B}-\tfrac{1}{c}\depp{}{t}\epsilon_0\vec{E}=\tfrac{1}{c}\vec{J} \\ \vec{\nabla}\cdot\tfrac{1}{\mu_0 c}\vec{B}=0 \\ \vec{\nabla}\times\epsilon_0\vec{E}+\tfrac{1}{c}\depp{}{t}\tfrac{1}{\mu_0 c}\vec{B}=0 \end{array}\right. \ ,
\end{equation*}with $c,\epsilon_0,\mu_0\in\CR$ real numbers satisfying $\epsilon_0\mu_0c^2=1$, $\rho$ and the vectors $\vec{J}$, $\vec{B}$, and $\vec{E}$ defined by 
\begin{equation*}
\boldsymbol{\mathrm{g}}\left(\rho\dep{}{t}+\Somatorio{j=1}{3}J_j\dep{}{x^j}, \ \cdot \ \right)=-c \ \alpha(\cdot) \ ,
\end{equation*} 
\begin{equation*}
\boldsymbol{\mathrm{g}}\left(\Somatorio{j=1}{3}B_j\dep{}{x^j}, \ \cdot \ \right)=-\mu_0c \ \omega\left(-\fraction{1}{c}\dep{}{t}, \ \cdot \ \right) \ ,
\end{equation*}and
\begin{equation*}
\boldsymbol{\mathrm{g}}\left(\Somatorio{j=1}{3}E_j\dep{}{x^j}, \ \cdot \ \right)=\fraction{1}{\epsilon_0}\star_{\boldsymbol{\mathrm{g}}}\omega\left(-\fraction{1}{c}\dep{}{t}, \ \cdot \ \right) \ .
\end{equation*}

An important fact to be noticed is that this split of the electromagnetic $2$-form into vectors $\vec{B}$ and $\vec{E}$ is unphysical and arbitrary, as it is not preserved by isometries ---it depends on a particular choice of inertial reference frame.

It is been implicitly assumed that I am describing Maxwell's theory on a background spacetime that is not dependent on the energy distribution associated to electromagnetic fields or their sources. It must be clearly stated: I am not dealing with the nonlinear coupled field equations of General Relativity and Electromagnetism in this note.

The classical dynamics of light and charged test particles are both described by hamiltonian flows. Light trajectories are null geodesics of $(M,\boldsymbol{\mathrm{g}})$, whilst the trajectories of a test particle, having mass $m\in\CR$ and charge $q\in\CR$, are integral curves of the hamiltonian flow defined by the function 
\begin{equation*}
T^*M\ni (v^*,x)\mapsto\tfrac{1}{2m}\boldsymbol{\mathrm{g}}\big{|}_{x}(mv,mv)\in\CR \ ,
\end{equation*}with $v^*(\cdot):=m\boldsymbol{\mathrm{g}}\big{|}_{x}(v,\cdot)\in T^*_xM$, on the cotangent bundle $T^*M$ endowed with the symplectic form $\ud\lambda+\mathrm{Pr}^*\left(\tfrac{q}{\epsilon_0 c}\star_{\boldsymbol{\mathrm{g}}}\omega\right)$, where $\ud\lambda$ is the canonical symplectic form of $T^*M$ induced by the projection $\mathrm{Pr}:T^*M\to M$. The Lorentz force law, 
\begin{equation*}
\vec{F}=\tfrac{q}{\epsilon_0}\left(\epsilon_0\vec{E}+\tfrac{\vec{v}}{c}\times\tfrac{1}{\mu_0 c}\vec{B}\right) \ ,
\end{equation*}can be deduced from this symplectic picture \cite{Sni1}. 


\section{The topological approach}

\hspace{1.5em}I shall consider only globally hyperbolic $4$-dimensional lorentzian manifolds with finite first and second Betti numbers to model spacetimes; in particular, a spacetime $M$ is diffeomorphic to $\mathcal{M}\times\CR$, and there exists a vector field $X\in\mathfrak{X}(M;\CR)$ which is the gradient (with respect to the lorentzian metric) of a global time function on $M$ (vide \cite{BerSan} and \cite{Wal}).

Let $e\in\CR$ be a real number, $J\subset M$ stand for the disjoint\footnote{Intricate circuits and collision between charged particles are not allowed, for the sake of simplicity.} union of worldlines and worldsheets corresponding to particle and wire distributions, and $M_J:=M-J$ be the complement of $J$. The next paragraph provides a precise definition for the electromagnetic field over a spacetime region that is compatible with Maxwell's theory, free of magnetic poles, and able to explain charge quantisation.

\vspace{1em}
\textit{Given charge and current distributions $J\subset M$, the electromagnetic field over a spacetime $(M,\boldsymbol{\mathrm{g}})$ is an equivalence class of hermitian line bundles with hermitian connexion over $M_J$, $[(L,\escalar{\cdot}{\cdot},\nabla^{\omega})]$, whose curvature $\omega:=-ie\cdot curv(\nabla^{\omega})$ of any representative satisfies $\integrall{N}{}{\star_{\boldsymbol{\mathrm{g}}}\omega}=0$, for every smooth singular $2$-cycle $N$.}
\vspace{1em}

Outside regions occupied by charges or currents, the electromagnetic $2$-form must satisfy the source free Maxwell's equations (if one's goal is a definition of an electromagnetic field compatible with Maxwell's theory),
\begin{equation*}
\left\{\begin{array}{c}\ud\omega=0 \\ \star_{\boldsymbol{\mathrm{g}}}\ud\star_{\boldsymbol{\mathrm{g}}}\omega=0 \end{array}\right. \ ,
\end{equation*}which is attained by the given definition. Indeed, the curvature of any hermitian connexion on a hermitian line bundle defines a closed differential $2$-form (cf. lemma \ref{unitarypotential2}), and a differential form whose integral over every smooth singular cycle vanishes is closed.

It is also not possible to probe inside these regions occupied by charges or currents, justifying the elimination of worldlines corresponding to particle (charge) distributions and worldsheets corresponding to wire (current) distributions from the spacetime region of interest. However, sources cannot be eliminated altogether; wherefore, one topologically introduces the source term back on Maxwell's equation via the following definitions. 

\vspace{1em}
 \textit{If the class $[N]$ is a generator of the second smooth singular homology group $H^\infty_2(M_J;\CZ)$, $\integrall{N}{}{\omega}$ is the charge inside a $3$-dimensional spacetime region bounded by any smooth singular $2$-cycle $N$ representing $[N]$; likewise, if the class $[\gamma]$ is a generator of the first smooth singular homology group $H^\infty_1(M_J;\CZ)$, the current passing through a $2$-dimensional spacetime region bounded by a smooth singular $1$-cycle $\gamma$ representing $[\gamma]$ is $-c\integrall{\gamma}{}{\imath_{X}\omega}$.} 
\vspace{1em}

It is important to mention that $\imath_{X}\omega$ is not closed unless $\pounds_X(\omega)=0$ (static fields), and this means that: if $\gamma\prime\in [\gamma]$, then $-c\integrall{\gamma}{}{\imath_{X}\omega}=-c\integrall{\gamma\prime}{}{\imath_{X}\omega}-c\integrall{N}{}{\pounds_X(\omega)}$ with $\partial N=\gamma-\gamma\prime$, and $-c\integrall{N}{}{\pounds_X(\omega)}$ is the displacement current. 

Due to the globally hyperbolic hypothesis imposed on $M$, a smooth singular $2$-cycle $N$ representing a generator $[N]\in H^\infty_2(M_J;\CZ)$ is exactly the same as a gaussian surface, and the above definition of the charge inside a region bounded by $N$ as the integral $\integrall{N}{}{\omega}$ is nothing more than the integral form of Gauss's law; as well as the condition $\integrall{N}{}{\star_{\boldsymbol{\mathrm{g}}}\omega}=0$ ---which guarantees the absence of magnetic poles. In the same manner, a smooth singular $1$-cycle $\gamma$ representing a generator $[\gamma]\in H^\infty_1(M_J;\CZ)$ corresponds to an amperian loop, and the definition of the current passing through the region bounded by it as $-c\integrall{\gamma}{}{\imath_{X}\omega}$ mimics Amp\`{e}re's circuital law, whilst Faraday's law is expressed by $\frac{1}{\epsilon_0}\integrall{\gamma}{}{\imath_{X}\star_{\boldsymbol{\mathrm{g}}}\omega}$ being the electromotive force. Indeed, $M\cong\mathcal{M}\times\CR$; thus, the four dimensional physical spacetime $M$ is homotopic equivalent to the three dimensional physical space $\mathcal{M}$, and smooth singular cycles generating the homology groups $H^\infty_1(M_J;\CZ)$ and $H^\infty_2(M_J;\CZ)$ can be taken from inside $\mathcal{M}$ minus the regions occupied by particle and wire distributions.

Equivalence classes of complex line bundles are classified by their first Chern class, and this is closely related to theorem \ref{teo1}: an equivalence class of hermitian line bundles over $M_J$ is completely determined by an element of $\check{H}^2(M_J;\CZ)$.

The first Chern class $c_1([L])\in\check{H}^2(M_J;\CZ)$ of $[(L,\escalar{\cdot}{\cdot},\nabla^{\omega})]$ is fixed by knowing the charges and quantal phases (cf. theorem \ref{teo3}). The periods of the curvature provide, via de Rham's theorem (cf. theorem \ref{teo5}), an element of the second de Rham cohomology group $[\frac{1}{e}\omega]\in H_{dR}^2(M_J;\CR)$ lying in the image of the homomorphism between $\check{H}^2(M_J;\CZ)$, the second \v{C}ech cohomology group with coefficients in the constant sheaf $\CZ$, and $H_{dR}^2(M_J;\CR)$, induced by the inclusion $\CZ\ni k\mapsto 2\pi k\in\CR$. Quantal phases\footnote{If the spacetime region has trivial fundamental group before taking out the worldsheets associated to current distributions, then the complex line bundle is chosen to be the one defined by the trivial homomorphism between $\pi_1(M_J)$ and $S^1$ ---there is no Aharonov--Bohm effects. When the spacetime region has nontrivial fundamental group before taking out the worldsheets associated to current distributions, one also needs to take this into account: besides the currents, one must know if there are electromagnetic fields inside the ``spacetime holes''.} provide a homomorphism between the fundamental group $\pi_1(M_J)$ and the unitary circle $S^1$, which defines an element on the kernel of the aforementioned homomorphism $\check{H}^2(M_J;\CZ)\rightarrow H_{dR}^2(M_J;\CR)$ (cf. lemma \ref{lem4} and theorem \ref{teo2}). 

The curvature is fixed by selecting a representative of $[\omega]\in H_{dR}^2(M_J;\CR)$ which satisfies $\integrall{N}{}{\star_{\boldsymbol{\mathrm{g}}}\omega}=0$ for every smooth singular $2$-cycle $N$, and matches the currents, and possible initial or boundary conditions (e.g. background electromagnetic radiation, or the vanishing of the electromagnetic $2$-form at some regions).

One can conclude from the above that magnetic effects are a manifestation of flat hermitian line bundles with nonflat connexion, whilst the Aharonov--Bohm effect is due to nontrivial flat hermitian line bundles with flat connexion; ergo, topologically, it is impossible to choose a different reference frame where an electric effect ---which is a manifestation of nonflat hermitian line bundles--- is seen as a magnetic effect. Furthermore, what is usually associated with the quantum aspects of electromagnetism is encoded in $\check{H}^2(M_J;\CZ)$ (Chern classes, quantisation of charge) and $\pi_1(M_J)$ (holonomy of flat connexions, Aharonov--Bohm effect), classical properties are due to $\Omega^2(M_J;\CR)$ (curvature, electromagnetic field tensor).

\begin{rema} The symplectic picture describing the classical dynamics of light and charged test particles, at the end of section \ref{PEMT}, remains true with $M$ replaced by $M_J$. The main difference is that undesired collisions between test particles and sources are avoided in $M_J$.
\end{rema}


\subsection{Topological charges and currents}

\hspace{1.5em}In order to demonstrate that these definitions imply quantisation of charges and do not support magnetic poles, I shall investigate the sources of electromagnetic fields on Minkowski spacetime.

First, $M\cong\mathcal{M}\times\CR$ with $\mathcal{M}\cong\CR^3$, and I am supposing that there is a particle\footnote{The particle need not to be a point particle: as long as it is homotopic equivalent to a point (e.g. a $3$-dimensional ball), it can have spatial extension or even be a distribution of particles.} inside the physical space $\mathcal{M}$ that may be a source of an electromagnetic field over the Minkowski spacetime $(M,\boldsymbol{\mathrm{g}})$.

In this particular example $J\subset M$ is the worldline of the particle (that may describe a complicated trajectory) and it is homotopic equivalent to a line. Consequently, $M_J$ is homotopic equivalent to $\CR^3-\{\text{point}\}$, and all possible inequivalent electromagnetic fields over the Minkowski spacetime $(M,\boldsymbol{\mathrm{g}})$ are parametrised by the integer numbers. Indeed, the fundamental group of $M_J$ is trivial, $\pi_1(M_J)=\{\boldsymbol{1}\}$; therefore, each equivalence class of hermitian line bundles with hermitian connexion over $M_J$, $[(L,\escalar{\cdot}{\cdot},\nabla^{\omega})]$, corresponds to an element of $\check{H}^2(M_J;\CZ)$ (cf. theorem \ref{teo3}), and this cohomology group is isomorphic to the second cohomology group of $\CR^3-\{\text{point}\}$ with integer coefficients, $H^2(\CR^3-\{\text{point}\};\CZ)$, which in turn is isomorphic to $\CZ$. 

Now I am going to assume that the electromagnetic field $[(L,\escalar{\cdot}{\cdot},\nabla^{\omega})]$ over the Minkowski spacetime $(M,\boldsymbol{\mathrm{g}})$ is the one corresponding to the integer $k\in\CZ$, and I want to show that not only the particle is the only possible source for this electromagnetic field, but it has an electric charge whose value is exactly $ke$. In order to do so, one must notice that the first smooth singular homology group $H^\infty_1(M_J;\CZ)$ is trivial and the second one, $H^\infty_2(M_J;\CZ)$, is isomorphic to $\CZ$ (they are simply the homology groups of $\CR^3-\{\text{point}\}$ with integer coefficients). This is equivalent to saying that all the amperian loops are trivial (they do not encompass any current other than displacement currents in nonstatic situations), and that there exist two types of gaussian surfaces: the ones that do not enclose the particle, and the ones that do. To take any smooth singular $2$-cycle $N$ representing a generator $[N]\in H^\infty_2(M_J;\CZ)$ is to take a gaussian surface enclosing the particle, and $\integrall{N}{}{\omega}=ke$ (cf. proposition \ref{prop1} and theorem \ref{teo5}). Wherefore, the electric charge is quantised and there are no other sources for this electromagnetic field, and no magnetic poles.

If instead of a single particle one considers $n$ particles (with nonintersecting worldlines), the reader can convince themselves that a Mayer--Vietoris argument guarantees that both $\check{H}^2(M_J;\CZ)$ and $H^\infty_2(M_J;\CZ)$ are isomorphic to $\CZ^n$, whilst both $\pi_1(M_J)$ and $H^\infty_1(M_J;\CZ)$ are trivial. For a gaussian surface $N$ enclosing all the particles, one has $\integrall{N}{}{\omega}=(k_1+\cdots+k_n)e$ with $k_1,\dots,k_n\in\CZ$ (cf. theorem \ref{teo5}) and the electromagnetic field $[(L,\escalar{\cdot}{\cdot},\nabla^{\omega})]$ over the Minkowski spacetime $(M,\boldsymbol{\mathrm{g}})$ being the tensor product of the electromagnetic fields corresponding to each integer $k_j$ (cf. theorem \ref{teo1}).

Finally, I want to address cases where wires are present. The first case is of an open wire and the second case is of a neutral closed wire; the wires need only to be homotopic equivalent to lines or circles, cylinders or solid tori (like copper wires) are typical examples. In both situations the electromagnetic field over the Minkowski spacetime is represented by flat hermitian line bundles with trivial holonomy group (for $\pi_1(M)=\{\boldsymbol{1}\}$, there cannot be Aharonov--Bohm effects).

After removing the worldsheet of an open wire from Minkowski spacetime, one has that $M_J$ is homotopic equivalent to $\CR^2-\{\text{point}\}$; hence both $\pi_1(M_J)$ and $H^\infty_1(M_J;\CZ)$ are isomorphic to $\CZ$, whilst $\check{H}^2(M_J;\CZ)$ and $H^\infty_2(M_J;\CZ)$ are trivial\footnote{Unfortunately, this approach cannot cope with the highly idealised situation of a charge distribution that cannot be enclosed in a finite volume spatial region, e.g. infinite long charged wires and plates.}. In similar fashion, by removing from the Minkowski spacetime the worldsheet of a closed wire, $M_J$ is now homotopic equivalent to $\CR^3-\{\text{circle}\}$; and as a result, $\pi_1(M_J)$ and $H^\infty_1(M_J;\CZ)$ are isomorphic to $\CZ$, whereas $\check{H}^2(M_J;\CZ)$ and $H^\infty_2(M_J;\CZ)$ are also isomorphic to $\CZ$ (the closed wire need not to be neutral, and its total charge is an integer multiple of $e$ given by $\integrall{N}{}{\omega}$, with $N$ any smooth singular $2$-cycle representing a generator $[N]\in H^\infty_2(M_J;\CZ)$). A smooth singular $1$-cycle $\gamma$ representing a generator $[\gamma]\in H^\infty_1(M_J;\CZ)$ is exactly an amperian loop around the wire in these situations; assuming that the wire is neutral, if it is carrying current, $\omega$ is different from zero, otherwise $\omega$ vanishes identically (the connexion is flat) ---there are no other sources for the electromagnetic $2$-form but the current passing through the wire.


\section{Semiclassical features}\label{SCf}

\hspace{1.5em}Let $(M,\boldsymbol{\mathrm{g}})=(\CR^4,\sum_{j=1}^{3}\ud x^j\otimes\ud x^j-\ud x^0\otimes\ud x^0)$ be the Minkowski spacetime with coordinates $(x^1,x^2,x^3,x^0=ct)$. Both $\pi_1(M)$ and $\check{H}^2(M;\CZ)$ are trivial; ergo, the trivial complex line bundle $L_0=\CC\times M$ is a representative for the source free electromagnetic field in vacuum. 

This section is devoted to unravel information contained in the sections of $L_0$.


\subsection{A wave function of a photon}

\hspace{1.5em}The trivial complex line bundle admits a flat connexion $\nabla^0$ satisfying 
\begin{equation*}
\nabla^0 fs=\ud f\otimes s\in\Omega^1(M;\CC)\otimes_{C^\infty(M;\CC)}\Gamma(L_0) \ ,
\end{equation*}for a unitary section of the complex line bundle $s\in\Gamma(L_0)$ and any complex-valued function $f\in C^\infty(M;\CC)$. 

There are two linear mappings, defined via the flat connexion, from $\mathfrak{X}(M;\CR)$ to the endomorphisms of $\Gamma(L_0)\oplus\Gamma(L_0)\oplus\Gamma(L_0)$ to be considered. Let $\hbar\in\CR$, taking $\omega_1,\omega_2,\omega_3\in\Gamma(L_0)$, $Y_j=\partial_j=\depp{}{x^j}\in\mathfrak{X}(M;\CR)$, and $X=\frac{-1}{c}\depp{}{t}\in\mathfrak{X}(M;\CR)$, the first mapping is given by: 
\begin{equation*}
Y_1\mapsto i\hbar\boldsymbol{\mathrm{y_1}} \ , \ \boldsymbol{\mathrm{y_1}}(\omega_1\oplus\omega_2\oplus\omega_3)=0\oplus i\nabla^0_{\partial_1}\omega_3\oplus -i\nabla^0_{\partial_1}\omega_2 \ , 
\end{equation*}
\begin{equation*}
Y_2\mapsto i\hbar\boldsymbol{\mathrm{y_2}} \ , \ \boldsymbol{\mathrm{y_2}}(\omega_1\oplus\omega_2\oplus\omega_3)=-i\nabla^0_{\partial_2}\omega_3\oplus 0\oplus i\nabla^0_{\partial_2}\omega_1 \ , 
\end{equation*}
\begin{equation*}
Y_3\mapsto i\hbar\boldsymbol{\mathrm{y_3}} \ , \ \boldsymbol{\mathrm{y_3}}(\omega_1\oplus\omega_2\oplus\omega_3)=i\nabla^0_{\partial_3}\omega_2\oplus -i\nabla^0_{\partial_3}\omega_1\oplus 0 \ , 
\end{equation*}and
\begin{equation*}
X\mapsto -i\hbar\boldsymbol{\mathrm{x}} \ , \ \boldsymbol{\mathrm{x}}(\omega_1\oplus\omega_2\oplus\omega_3)=-(\nabla^0_X\omega_1\oplus\nabla^0_X\omega_2\oplus\nabla^0_X\omega_3) \ .
\end{equation*}The second mapping is defined by
\begin{equation*}
Y_1\mapsto i\hbar\boldsymbol{\mathrm{x_1}} \ , \ \boldsymbol{\mathrm{x_1}}(\omega_1\oplus\omega_2\oplus\omega_3)=\nabla^0_{\partial_1}\omega_1+\nabla^0_{\partial_2}\omega_2+\nabla^0_{\partial_3}\omega_3\oplus 0\oplus 0 \ ,
\end{equation*}
\begin{equation*}
Y_2\mapsto i\hbar\boldsymbol{\mathrm{x_2}} \ , \ \boldsymbol{\mathrm{x_2}}(\omega_1\oplus\omega_2\oplus\omega_3)=0\oplus\nabla^0_{\partial_1}\omega_1+\nabla^0_{\partial_2}\omega_2+\nabla^0_{\partial_3}\omega_3\oplus 0 \ , 
\end{equation*}
\begin{equation*}
Y_3\mapsto i\hbar\boldsymbol{\mathrm{x_3}} \ , \ \boldsymbol{\mathrm{x_3}}(\omega_1\oplus\omega_2\oplus\omega_3)=0\oplus 0\oplus\nabla^0_{\partial_1}\omega_1+\nabla^0_{\partial_2}\omega_2+\nabla^0_{\partial_3}\omega_3 \ , 
\end{equation*}and
\begin{equation*}
X\mapsto -i\hbar\boldsymbol{\mathrm{x_0}} \ , \ \boldsymbol{\mathrm{x_0}}(\omega_1\oplus\omega_2\oplus\omega_3)=0\oplus 0\oplus 0 \ .
\end{equation*}

Let $L_0^{-1}$ denote the dual bundle of $L_0$. A hermitian structure $\escalar{\cdot}{\cdot}$ provides an equivalence between the dual bundle and the complex conjugate bundle $\overline{L}_0$ (defined by the complex conjugate of the transition functions of $L_0$). In the particular case of trivial complex line bundles, $L_0\cong L_0^{-1}$. 

I am now in a position to furnish a definition of a wave function for a photon that agrees with previous ones introduced in \cite{Oppa} and \cite{Bir}.

\vspace{1em}
\textit{Wave functions for photons are sections of $\Gamma(L_0)\oplus\Gamma(L_0)\oplus\Gamma(L_0)\oplus\Gamma\left(\overline{L}_0\right)\oplus\Gamma\left(\overline{L}_0\right)
\oplus\Gamma\left(\overline{L}_0\right)$ of the form $\omega_1^{+}\oplus\omega_2^{+}\oplus\omega_3^{+}\oplus\omega_1^{-}\oplus\omega_2^{-}\oplus\omega_3^{-}$, with $\omega_1^{+}\oplus\omega_2^{+}\oplus\omega_3^{+}\in\Gamma(L_0)\oplus\Gamma(L_0)\oplus\Gamma(L_0)$ lying in the intersection of the kernels of the endomorphisms induced by $\sum_{j=1}^{3}Y_j+X$,} 
\begin{equation*}
\left\{\begin{array}{l}
i\hbar\left(\Somatorio{j=1}{3}\boldsymbol{\mathrm{y_j}}-\boldsymbol{\mathrm{x}}\right)\omega_1^{+}\oplus\omega_2^{+}\oplus\omega_3^{+}=0 \\
i\hbar\left(\Somatorio{j=1}{3}\boldsymbol{\mathrm{x_j}}-\boldsymbol{\mathrm{x_0}}\right)\omega_1^{+}\oplus\omega_2^{+}\oplus\omega_3^{+}=0 
\end{array}\right. \ ,
\end{equation*}\textit{and $\omega_1^{-}\oplus\omega_2^{-}\oplus\omega_3^{-}\in\Gamma\left(\overline{L}_0\right)\oplus\Gamma\left(\overline{L}_0\right)
\oplus\Gamma\left(\overline{L}_0\right)$, under the identification $L_0\cong\overline{L}_0$, satisfying} 
\begin{equation*}
\left\{\begin{array}{l}
i\hbar\left(-\Somatorio{j=1}{3}\boldsymbol{\mathrm{y_j}}-\boldsymbol{\mathrm{x}}\right)\omega_1^{-}\oplus\omega_2^{-}\oplus\omega_3^{-}=0 \\
i\hbar\left(\Somatorio{j=1}{3}\boldsymbol{\mathrm{x_j}}-\boldsymbol{\mathrm{x_0}}\right)\omega_1^{-}\oplus\omega_2^{-}\oplus\omega_3^{-}=0 
\end{array}\right. \ ,
\end{equation*}\textit{together with a compatibility condition, considering $L_0\cong L_0^{-1}$,}
\begin{equation*}
\escalar{\omega_j^{-}}{\cdot}=\omega_j^{+}(\cdot) \ .
\end{equation*}
\vspace{1em}

Using the unitary section $s$, sections of the trivial complex line bundle can be identified with complex-valued functions, $\omega_j^{+}=\left(\tfrac{1}{e\mu_0c}B_j+i\tfrac{\epsilon_0}{e}E_j\right)s$, and one recovers a wave function for a photon (having a particular helicity) as in references \cite{Oppa} and \cite{Bir}. 


\subsection{Dirac's equation}   

\hspace{1.5em}Considering now a connexion $\nabla^\star$ in $L_0$ whose curvature is $\tfrac{ie}{\epsilon_0c\hbar}\star_{\boldsymbol{\mathrm{g}}}\omega$, one has a linear mapping from $\mathfrak{X}(M;\CR)$ to the endomorphisms of $\Gamma(L_0)\oplus\Gamma(L_0)\oplus\Gamma\left(\overline{L}_0\right)\oplus\Gamma\left(\overline{L}_0\right)$ defined by $Y_j\mapsto i\hbar\boldsymbol{\mathrm{y_j}}(\omega)$ and $X\mapsto-i\hbar\boldsymbol{\mathrm{x}}(\omega)$: if $\Psi_1,\Psi_2,\Psi_3,\Psi_4\in\Gamma(L_0)$, and considering the identification $L_0\cong\overline{L}_0$, then 
\begin{equation*}
\boldsymbol{\mathrm{y_1}}(\omega)(\Psi_1\oplus\Psi_2\oplus\Psi_3\oplus\Psi_4)=\nabla^\star_{\partial_1}\Psi_4\oplus\nabla^\star_{\partial_1}\Psi_3\oplus-\nabla^\star_{\partial_1}
\Psi_2\oplus-\nabla^\star_{\partial_1}\Psi_1 \ , 
\end{equation*}\begin{equation*}
\boldsymbol{\mathrm{y_2}}(\omega)(\Psi_1\oplus\Psi_2\oplus\Psi_3\oplus\Psi_4)=-i\nabla^\star_{\partial_2}\Psi_4\oplus i\nabla^\star_{\partial_2}\Psi_3\oplus i\nabla^\star_{\partial_2}\Psi_2\oplus-i\nabla^\star_{\partial_2}\Psi_1 \ , 
\end{equation*}\begin{equation*}
\boldsymbol{\mathrm{y_3}}(\omega)(\Psi_1\oplus\Psi_2\oplus\Psi_3\oplus\Psi_4)=\nabla^\star_{\partial_3}\Psi_3\oplus-\nabla^\star_{\partial_3}\Psi_4\oplus-\nabla^\star_{\partial_3}
\Psi_1\oplus\nabla^\star_{\partial_3}\Psi_2 \ ,
\end{equation*}and\begin{equation*}
\boldsymbol{\mathrm{x}}(\omega)(\Psi_1\oplus\Psi_2\oplus\Psi_3\oplus\Psi_4)=-\nabla^\star_X\Psi_1\oplus-\nabla^\star_X\Psi_2\oplus\nabla^\star_X\Psi_3\oplus\nabla^\star_X\Psi_4 \ .
\end{equation*}

\vspace{1em}
\textit{Wave functions describing electrons and positrons of mass $m\in\CR$ subjected to an electromagnetic field $[(L,\escalar{\cdot}{\cdot},\nabla^{\omega})]$ are eigenvectors of the endomorphism induced by $\sum_{j=1}^{3}Y_j+X$ whose eigenvalue is $mc$:}
\begin{equation*}
i\hbar\left(\Somatorio{j=1}{3}\boldsymbol{\mathrm{y_j}}(\omega)-\boldsymbol{\mathrm{x}}(\omega)\right)\Psi_1\oplus\Psi_2\oplus\Psi_3\oplus\Psi_4=mc \ \Psi_1\oplus\Psi_2\oplus\Psi_3\oplus\Psi_4 \ .
\end{equation*}
\vspace{1em} 

This yields a semiclassical treatment for test charged particles under the influence of a fixed background electromagnetic field. Again, by using the unitary section $s$, for each $\Psi_j$ there exists a complex-valued function $\psi_j\in C^\infty(M;\CC)$ such that $\Psi_j=\psi_j s$, and one recovers Dirac's equation for a wave function for an electron and a positron \cite{PaulDirac28}, justifying the previous definition. 

\begin{rema} The whole analysis of section \ref{SCf} applies verbatim even for globally hyperbolic spacetimes which are not diffeomorphic to $\CR^4$. The trivial complex line bundle is still a representative for the source free electromagnetic field in vacuum, the class of $\tfrac{e}{\epsilon_0c\hbar}\star_{\boldsymbol{\mathrm{g}}}\omega$ is still trivial (independent of the fact that $\check{H}^2(M;\CZ)$ might not be), and there are well defined vector fields $Y_1,Y_2,Y_3,X\in\mathfrak{X}(M;\CR)$ which are linearly independent at every point of $M$. This last nontrivial fact is consequence of the globally hyperbolic hypothesis imposed on the spacetime $M$. Since $M\cong\mathcal{M}\times\CR$ and the orientability of it imply that $\mathcal{M}$ is orientable, one can apply Whitehead's theorem \cite{Wh} to conclude that the tangent bundle of $M$ is trivial. If one wants to consider the Aharonov-Bohm effect in this formalism, instead of $L_0$ one uses $L_0\otimes L_{\star}$, where $[L_{\star}]$ is a nontrivial class of complex line bundles defined by the de Rham class of $\frac{e}{\epsilon_0c\hbar}\star_{\boldsymbol{\mathrm{g}}}\omega$ and the quantal phases. 
\end{rema}


\setcounter{section}{1}

\renewcommand{\thesection}{\Alph{section}}

\addcontentsline{toc}{section}{Appendix: Hermitian line bundles}

\section*{Appendix: Hermitian line bundles}

\hspace{1.5em}This is a summary of relevant results and definitions needed to understand this note (although technicalities will be avoided here), the reader can find them in references \cite{BrD,Ko,Lee}. 

I shall assume that the cohomology and homology groups discussed in this appendix have finite dimension.

\begin{defi} A complex vector bundle $\mathscr{E}$ of rank $k$ over a manifold $M$ is a manifold together with a smooth surjective mapping $\mathscr{F}:\mathscr{E}\to M$ such that: for any point $p\in M$ there exist a neighbourhood $A\subset M$ containing it and a diffeomorphism $\varphi_A:\mathscr{F}^{-1}(A)\to \CC^k\times A$, whose restriction $\varphi_A\big{|}_{\mathscr{F}^{-1}(q)}:\mathscr{F}^{-1}(q)\to\CC^k\times\{q\}$ is a $\CC$-linear isomorphism at every $q\in A$, satisfying $\mathscr{F}\big{|}_{\mathscr{F}^{-1}(A)}=\mathrm{pr_A}\circ\varphi_A$, with $\mathrm{pr_A}:\CC^k\times A\to A$ being the projection onto the second factor.
\end{defi}

A neighbourhood such as $A$ is called a trivialising neighbourhood, and, when the surjective mapping $\mathscr{F}$ is omitted, $\mathscr{E}\big{|}_{A}:=\mathscr{F}^{-1}(A)$ represents the restriction of the bundle $\mathscr{E}$ on the neighbourhood $A\subset M$ and $\mathscr{E}_p:=\mathscr{F}^{-1}(p)\cong\CC^k$ stands for the fibre over $p\in M$.

\begin{defi} A smooth section of a complex vector bundle $\mathscr{F}:\mathscr{E}\stackrel{\CC^k}{\longrightarrow} M$ is a smooth mapping $s:M\to\mathscr{E}$ satisfying $\mathscr{F}\circ s(p)=p$ for all $p\in M$. 
\end{defi}

For each open set $A\subset M$, a local section of a complex vector bundle is a section of $\mathscr{E}\big{|}_{A}$. One can, then, define the sheaf of sections of a complex vector bundle.

\begin{defi} The sheaf of smooth sections of a complex vector bundle $\mathscr{E}$ is denoted by $\mathcal{S}$, and $\Gamma(\mathscr{E})$ is the $C^\infty(M;\CC)$-module of global smooth sections of the complex vector bundle. 
\end{defi}

The $k$-th \v{C}ech cohomology group of a manifold $M$ with coefficients in a sheaf $\mathcal{Z}$ will be denoted by $\check{H}^k(M;\mathcal{Z})$, in particular $\Gamma(\mathscr{E})\cong\check{H}^0(M;\mathcal{S})$ as vector spaces.

\begin{defi} A hermitian line bundle $L\stackrel{\CC}{\longrightarrow} M$ is a complex vector bundle over $M$ with fibres isomorphic to $\CC$, together with a hermitian structure $\escalar{\cdot}{\cdot}$ defined on its space of sections $\Gamma(L)$.
\end{defi}

\begin{prop} For hermitian line bundles, the existence of a trivialisation is equivalent to the existence of a unitary section.
\end{prop}
\textit{Proof:} Let $A_j$ be a trivialising neighbourhood of a hermitian line bundle $(L,\escalar{\cdot}{\cdot})$. There exists a diffeomorphism $\varphi_j:L\big{|}_{A_j}\to\CC\times A_j$, and if $z\in\CC$ belongs to the unitary circle, then one can define a section of $\CC\times A_j$ via $A_j\ni p\mapsto (z,p)$. A unitary section $s_j\in\Gamma(L\big{|}_{A_j})$ is, then, obtained by $s_j(p)=\varphi_j^{-1}(z,p)$.

Conversely, let $s_j\in\Gamma(L\big{|}_{A_j})$ be a section satisfying $\escalar{s_j}{s_j}(p)=1$ for all $p\in A_j$. The mapping given by $\CC\times A_j\ni (z,p)\mapsto z\cdot s_j(p)\in L\big{|}_{A_j}$ is a diffeomorphism.\hfill $\blacksquare$\par\vspace{0.5em}
 
\begin{prop} Sections of $L$ can be represented by complex-valued functions over trivialising neighbourhoods.
\end{prop}
\textit{Proof:} Let $A_j$ be a trivialising neighbourhood of $L$ with unitary section $s_j$. The complex line bundle $L\big{|}_{A_j}\stackrel{\CC}{\longrightarrow}A_j$ is diffeomorphic to the trivial one, $\CC\times A_j$, and under this diffeomorphism, a section $s\in\Gamma(L\big{|}_{A_j})$ is uniquely defined by a complex-valued function $f\in C^\infty(A_j;\CC)$, i.e. $s(p)\simeq(f(p),p)$ for each point $p\in A_j$.\hfill $\blacksquare$\par\vspace{0.5em}

When there is an identification between a section $s$ of a complex line bundle and a complex-valued function $f$, the diffeomorphism might be omitted, for the sake of simplicity, and the equality $s=f$ will be used.

Two hermitian line bundles $(L,\escalar{\cdot}{\cdot})$ and $(L',\escalar{\cdot}{\cdot}')$ over the same manifold $M$ are equivalent if there exists a diffeomorphism $\phi:L\to L'$ inducing a $C^\infty(M;\CC)$-module isomorphism between $\Gamma(L)$ and $\Gamma(L')$ via $\Gamma(L)\ni s\mapsto\phi\circ s\in\Gamma(L')$ satisfying $\escalar{\phi\circ r}{\phi\circ s}'=\escalar{r}{s}$ for any pair of sections $r,s\in\Gamma(L)$. Such a diffeomorphism will be called a hermitian bundle diffeomorphism. In particular, when restricted to the zero section $M\subset L$, $\phi\big{|}_M:M\subset L\to L'$ is the identity on $M$ and, when restricted to a fibre $L_p$, $\phi\big{|}_{L_p}:L_p\subset L\to L'$ is an isometry between $L_p$ and ${L'}_p$. 

\begin{defi}The set of equivalence classes of hermitian line bundles endowed with the group structure given by tensor products of hermitian line bundles will be denoted by $\boldsymbol{\mathcal{L}_h}(M)$.
\end{defi}

Let $\mathcal{C}^\infty_{\CR}$ and $\mathcal{C}^\infty_{S^1}$ stand for the sheaves of functions over a manifold $M$ with values on $\CR$ and $S^1$. It holds that $\check{H}^0(M;\mathcal{C}^\infty_{\CR})=C^\infty(M;\CR)$ and $\check{H}^0(M;\mathcal{C}^\infty_{S^1})=C^\infty(M;S^1)$. 

\begin{lem}\label{lem1}As groups, $\boldsymbol{\mathcal{L}_h}(M)\cong\check{H}^1(M;\mathcal{C}^\infty_{S^1})$.
\end{lem}
\textit{Proof:} Given a contractible open cover $\mathcal{A}=\{A_j\}_{j\in I}$ of $M$ (which can always be obtained by using a convenient cover made of balls with respect to a Riemannian metric) and an element $\varphi$ of $\check{H}^1(M;\mathcal{C}^\infty_{S^1})$, one can define a hermitian line bundle $L$ such that each $A_j$ is a trivialising neighbourhood of it with transition functions given by $\varphi(A_j\cap A_k)$. 

The construction goes as follows: one can define $L\big{|}_{A_j}:=\CC\times A_j$ with the standard hermitian structure (or any other one) on the fibre $\CC$ and consider the equivalence relation $(z,p)\sim (\varphi(A_j\cap A_k)(p)\cdot z,p)$ over $A_j\cap A_k$; then, the set $L:=\displaystyle\bigsqcup_{j\in I}L\big{|}_{A_j}/\sim$ has both a natural smooth structure, making it a complex line bundle over $M$, and a natural hermitian structure $\escalar{\cdot}{\cdot}$.

Now, supposing that the hermitian line bundle $(L,\escalar{\cdot}{\cdot})$ is given together with a contractible open cover $\mathcal{A}=\{A_j\}_{j\in I}$ of $M$ where each $A_j$ is a trivialising neighbourhood, i.e. there exist bundle morphisms $\varphi_j:L\big{|}_{A_j}\to\CC\times A_j$ respecting the hermitian structure, one can define an element of $\check{H}^1(M;\mathcal{C}^\infty_{S^1})$. The diffeomorphisms
\begin{equation*}
 \varphi_k\big{|}_{A_j\cap A_k}\circ\varphi_j\big{|}_{A_j\cap A_k}:\CC\times A_j\cap A_k\to \CC\times A_j\cap A_k
\end{equation*}induce mappings satisfying the cocycle conditions; the transition functions 
\begin{equation*}
\varphi(A_j\cap A_k):A_j\cap A_k\to S^1
\end{equation*}defined by 
\begin{equation*}
\varphi_k\big{|}_{A_j\cap A_k}\circ\varphi_j\big{|}_{A_j\cap A_k}(z,p)=(\varphi(A_j\cap A_k)(p)\cdot z,p) \ .
\end{equation*}\hfill $\blacksquare$\par\vspace{0.5em}

\begin{lem}\label{lem2}As groups, $\check{H}^1(M;\mathcal{C}^\infty_{S^1})\cong\check{H}^2(M;\CZ)$.
\end{lem}
\textit{Proof:} The sequence
\begin{equation*}
0\hookrightarrow\CZ\hookrightarrow\mathcal{C}^\infty_{\CR}\twoheadrightarrow\mathcal{C}^\infty_{S^1}\twoheadrightarrow 1
\end{equation*}
\begin{equation*}
0\mapsto 0 \ , \ k\mapsto 2\pi k \ , \ f\mapsto \e^{if} \ , \ \varphi\mapsto 1
\end{equation*}is a short exact sequence of sheaves. Indeed, 
\begin{equation*}
\mathrm{ker}\{f\mapsto \e^{if}\}=2\pi\CZ=\mathrm{im}\{k\mapsto 2\pi k\} \ .
\end{equation*}Thus, there is a long exact sequence of cohomology groups, and the fine property of the sheaf $\mathcal{C}^\infty_{\CR}$ provides the result.\hfill $\blacksquare$\par\vspace{0.5em}

\begin{teo}\label{teo1}As groups, $\boldsymbol{\mathcal{L}_h}(M)\cong\check{H}^2(M;\CZ)$.
\end{teo}
\textit{Proof:} Lemmata \ref{lem1} and \ref{lem2}.\hfill $\blacksquare$\par\vspace{0.5em}

\begin{prop} $\check{H}^1(M;S^1)\hookrightarrow\check{H}^1(M;\mathcal{C}^\infty_{S^1})$.
\end{prop}
\textit{Proof:} Each element $\varphi\in\check{H}^1(M;S^1)$ provides a mapping 
\begin{equation*}
A_j\cap A_k\ni p\mapsto\varphi_{jk}(p):=\varphi(A_j\cap A_k)\in S^1
\end{equation*}satisfying the cocycle conditions. \hfill $\blacksquare$\par\vspace{0.5em}

Under pertinent identifications, hermitian line bundles defined by elements of $\check{H}^1(M;S^1)$ are called flat.

\begin{teo}\label{teo2}The kernel of the homomorphism $\check{H}^2(M;\CZ)\to\check{H}^2(M;\CR)$ induced by the inclusion $\CZ\hookrightarrow\CR$ sits in the following short exact sequence of groups: 
\begin{equation*}
0\hookrightarrow\fraction{\check{H}^1(M;\CR)}{\check{H}^1(M;\CZ)}\hookrightarrow\check{H}^1(M;S^1)\twoheadrightarrow\mathrm{ker}\{\check{H}^2(M;\CZ)\to\check{H}^2(M;\CR)\} \twoheadrightarrow 0 \ .
\end{equation*}
\end{teo}
\textit{Proof:} The sequence
\begin{equation*}
0\hookrightarrow\CZ\hookrightarrow\CR\twoheadrightarrow S^1\twoheadrightarrow 1
\end{equation*}
\begin{equation*}
0\mapsto 0 \ , \ k\mapsto 2\pi k \ , \ a\mapsto \e^{ia} \ , \ \varphi\mapsto 1
\end{equation*}is a short exact sequence of sheaves: 
\begin{equation*}
\mathrm{ker}\{a\mapsto \e^{ia}\}=2\pi\CZ=\mathrm{im}\{k\mapsto 2\pi k\} \ .
\end{equation*}The long exact sequence of cohomology groups associated with it is
\begin{equation*}
0\hookrightarrow\check{H}^1(M;\CZ)\hookrightarrow\check{H}^1(M;\CR)\rightarrow\check{H}^1(M;S^1)\rightarrow\check{H}^2(M;\CZ)\rightarrow\check{H}^2(M;\CR) \ ;
\end{equation*}then, 
\begin{equation*}
\mathrm{im}\{\check{H}^1(M;S^1)\to\check{H}^2(M;\CZ)\}\cong\mathrm{ker}\{\check{H}^2(M;\CZ)\to\check{H}^2(M;\CR)\} \ ,
\end{equation*}and the first isomorphism theorem implies that 
\begin{equation*}
\mathrm{ker}\{\check{H}^1(M;S^1)\to\check{H}^2(M;\CZ)\}\cong\fraction{\check{H}^1(M;\CR)}{\check{H}^1(M;\CZ)} \ .
\end{equation*}\hfill $\blacksquare$\par\vspace{0.5em}

Using an isomorphism between \v{C}ech and de Rham cohomology, a class on the image of the homomorphism $\check{H}^2(M;\CZ)\to H^2_{dR}(M;\CR)\cong\check{H}^2(M;\CR)$ induced by the inclusion $\CZ\hookrightarrow\CR$ are said to be integral. 

\begin{lem}\label{lem3} Let $K^2(\CZ;\CR)$ be the kernel and $I^2(\CZ;\CR)$ the image of the homomorphism $\check{H}^2(M;\CZ)\to\check{H}^2(M;\CR)$ induced by the inclusion $\CZ\hookrightarrow\CR$. The exact sequence 
\begin{equation*}
0\hookrightarrow K^2(\CZ;\CR)\hookrightarrow\check{H}^2(M;\CZ)\twoheadrightarrow I^2(\CZ;\CR)\twoheadrightarrow 0
\end{equation*}splits.
\end{lem}
\textit{Proof:} By the first isomorphism theorem, 
\begin{equation*}
\mathrm{im}\{\check{H}^2(M;\CZ)\to\check{H}^2(M;\CR)\}\cong \fraction{\check{H}^2(M;\CZ)}{\mathrm{ker}\{\check{H}^2(M;\CZ)\to\check{H}^2(M;\CR)\}} \ ,
\end{equation*}and, under this identification, the mapping $\check{H}^2(M;\CZ)\twoheadrightarrow I^2(\CZ;\CR)$ takes an element of $\check{H}^2(M;\CZ)$ to its class in $\fraction{\check{H}^2(M;\CZ)}{K^2(\CZ;\CR)}$. A right split is defined by taking a basis of $\fraction{\check{H}^2(M;\CZ)}{K^2(\CZ;\CR)}$ and choosing an arbitrary representative in $\check{H}^2(M;\CZ)$ for each basis element. \hfill $\blacksquare$\par\vspace{0.5em}

\begin{lem}\label{lem4}Let $\mathrm{Hom}(\pi_1(M);S^1)$ be the group formed by the group homomorphisms between the fundamental group of $M$, $\pi_1(M)$, and the unitary circle $S^1$. As groups, $\check{H}^1(M;S^1)\cong\mathrm{Hom}(\pi_1(M);S^1)$.  
\end{lem}
\textit{Proof:} \v{C}ech and singular cohomologies are isomorphic under the topological assumptions on $M$, in particular, $\check{H}^1(M;S^1)\cong H^1(M;S^1)$, and it suffices to prove $H^1(M;S^1)\cong\mathrm{Hom}(\pi_1(M);S^1)$. The isomorphism is provided by the following construction: a $1$-cocycle representing a generator of $H^1(M;S^1)$ can be understood as a homomorphism (respecting the abelian structures) between the group of singular $1$-chains over $M$ and $S^1$, and those homomorphisms can be mapped to elements of $\mathrm{Hom}(\pi_1(M);S^1)$, for a generator of $\pi_1(M)$ is represented by a singular $1$-chain.\hfill $\blacksquare$\par\vspace{0.5em}

\begin{teo}\label{teo3} One can uniquely define an equivalence class of hermitian line \linebreak bundles from a mapping belonging to $\mathrm{Hom}(\pi_1(M);S^1)$ and a de Rham class $[\omega]\in H_{dR}^2(M;\CR)$, with $[\frac{1}{e}\omega]$ integral for some $e\in\CR$.
\end{teo}
\textit{Proof:} Theorem \ref{teo1} and lemma \ref{lem3} imply $\boldsymbol{\mathcal{L}_h}(M)\cong I^2(\CZ;\CR)\oplus K^2(\CZ;\CR)$, furthermore theorem \ref{teo2} and lemma \ref{lem4} guarantee that an element of $K^2(\CZ;\CR)$ is determined by an element of $\mathrm{Hom}(\pi_1(M);S^1)$, and the identification $\check{H}^2(M;\CR)\cong H^2_{dR}(M;\CR)$ guarantees that an integral class $[\frac{1}{e}\omega]$ determines an element of $I^2(\CZ;\CR)$. \hfill $\blacksquare$\par\vspace{0.5em}

\begin{defi} A hermitian connexion on a hermitian line bundle $(L,\escalar{\cdot}{\cdot})$ over a manifold $M$ is a linear mapping 
\begin{equation*}
\nabla:\Gamma(L)\to\Omega^1(M;\CC)\otimes_{C^\infty(M;\CC)}\Gamma(L)
\end{equation*}satisfying: for any $r,s\in\Gamma(L)$, $f\in C^\infty(M;\CC)$ and $X\in\mathfrak{X}(M;\CR)$,
\begin{equation*}
\nabla(fs)=\ud f\otimes s +f\nabla s \ , 
\end{equation*}with $(\nabla s)(X)$ denoted by $\nabla_X s$, and
\begin{equation*}
X(\escalar{r}{s})=\escalar{\nabla_X r}{s}+\escalar{r}{\nabla_X s} \ . 
\end{equation*}
\end{defi}

With respect to a unitary section $s_j$, defined over a trivialising neighbourhood $A_j\subset M$, the connexion $\nabla$ can be represented by a potential $1$-form $\Theta_j\in\Omega^1(A_j;\CR)$: 
\begin{equation*}
\nabla s_j=i\Theta_j\otimes s_j \ . 
\end{equation*}Indeed, since $s_j$ is unitary, $\escalar{s_j}{s_j}=1$, for any $X\in\mathfrak{X}(A_j;\CR)$ it holds
\begin{align*}
0=X(\escalar{s_j}{s_j})&=\escalar{\nabla_X s_j}{s_j}+\escalar{s_j}{\nabla_X s_j}=\escalar{i\Theta_j(X)s_j}{s_j}+\escalar{s_j}{i\Theta_j(X)s_j} \nonumber \\
&=-i\overline{\Theta_j(X)}\escalar{s_j}{s_j}+i\Theta_j(X)\escalar{s_j}{s_j}=-i\overline{\Theta_j(X)}+i\Theta_j(X) \ .
\end{align*}

\begin{lem}\label{unitarypotential}The potential $1$-forms, $\Theta_j$ and $\Theta_k$, of $\nabla$ for each unitary section, $s_j$ and $s_k$, defined over an intersection $A_j\cap A_k$ are cohomologous. 
\end{lem}
\textit{Proof:} Since $L$ is a hermitian line bundle, there exists a transition function \newline $\e^{if_{jk}}\in C^\infty(A_j\cap A_k;S^1)$ relating the two unitary sections $s_j=\e^{if_{jk}}s_k$; as a result,
\begin{equation*}
\nabla s_j=i\Theta_j\otimes s_j=i\e^{if_{jk}}\Theta_j\otimes s_k \ , 
\end{equation*}however 
\begin{equation*}
\nabla s_j=\nabla (\e^{if_{jk}}s_k)=(\ud\e^{if_{jk}}+i\e^{if_{jk}}\Theta_k)\otimes s_k=i\e^{if_{jk}}(\ud f_{jk}+\Theta_k)\otimes s_k
\end{equation*}and, therefore,
\begin{equation*}
i\e^{if_{jk}}\Theta_j\otimes s_k=i\e^{if_{jk}}(\ud f_{jk}+\Theta_k)\otimes s_k \ \Rightarrow 
\end{equation*}
\begin{equation*}
\Theta_j-\Theta_k=\ud f_{jk} \ . 
\end{equation*}\hfill $\blacksquare$\par\vspace{0.5em}

\begin{defi}
The curvature of a hermitian connexion $\nabla$ on a hermitian line bundle $(L,\escalar{\cdot}{\cdot})$ over a manifold $M$ is the imaginary $2$-form $curv(\nabla)\in\Omega^2(M;i\CR)$ satisfying, for any $s\in\Gamma(L)$ and $X,Y\in\mathfrak{X}(M;\CR)$, 
 \begin{equation*}
curv(\nabla)(X,Y)s=\nabla_X\circ\nabla_Y s-\nabla_Y\circ\nabla_X s-\nabla_{[X,Y]}s \ .
\end{equation*}
\end{defi}

The next lemma implies, in particular, that the curvature is closed. 

\begin{lem}\label{unitarypotential2}If $\Theta$ is a potential $1$-form for the connexion $\nabla$, over a trivialising neighbourhood $A\subset M$, then its curvature is given by $curv(\nabla)=i\ud\Theta$, and it is independent of the choice of trivialisation. 
\end{lem}
\textit{Proof:} Assuming that $s$ is the unitary section associated with $\Theta$, for any $X,Y\in\mathfrak{X}(A;\CR)$, 
\begin{align*}
curv(\nabla)(X,Y)s&=\nabla_X\circ\nabla_Ys-\nabla_Y\circ\nabla_Xs-\nabla_{[X,Y]}s \nonumber \\
&=\nabla_X(i\Theta(Y)s)-\nabla_Y(i\Theta(X)s)-i\Theta([X,Y])s \nonumber \\
&=i\ud(\imath_Y\Theta)(X)s-\Theta(Y)\Theta(X)s \nonumber \\ 
&\quad-i\ud(\imath_X\Theta)(Y)s+\Theta(X)\Theta(Y)s-i\Theta([X,Y])s \nonumber \\
&=i(\ud(\imath_Y\Theta)(X)-\ud(\imath_X\Theta)(Y)-\Theta([X,Y]))s \nonumber \\ 
&=i(\ud\Theta)(X,Y)s \ .
\end{align*}Independence of the choice of trivialisation is provided by lemma \ref{unitarypotential}.\hfill $\blacksquare$\par\vspace{0.5em}

Two equivalent hermitian line bundles $(L,\escalar{\cdot}{\cdot})$ and $(L',\escalar{\cdot}{\cdot}')$ over the same manifold $M$ have equivalent hemitian connexions $\nabla$ and $\nabla'$ if there exists a hermitian bundle diffeomorphism $\phi:L\to L'$ satisfying $\phi\circ\nabla_Xs={\nabla'}_{X}\phi\circ s$ for any section $s\in\Gamma(L)$ and vector field $X\in\mathfrak{X}(M;\CR)$.

\begin{lem}The curvatures of two equivalent hermitian line bundles with hermitian connexions $(L,\escalar{\cdot}{\cdot},\nabla)$ and $(L',\escalar{\cdot}{\cdot}',\nabla')$ are equal.
\end{lem}
\textit{Proof:} Locally, the condition 
\begin{equation*}
\phi\circ\nabla_Xs={\nabla'}_{X}\phi\circ s
\end{equation*}implies that the potential $1$-forms are equal.\hfill $\blacksquare$\par\vspace{0.5em}

\begin{teo}\label{teo4}
For a fixed closed form $\omega\in\Omega^2(M;\CR)$ whose de Rham class $[\frac{1}{e}\omega]$ is integral for a real number $e\in\CR$, there exists an equivalence class of hermitian line bundles with hermitian connexion $[(L,\escalar{\cdot}{\cdot},\nabla^{\omega})]$ such that $-ie\cdot curv(\nabla^{\omega})=\omega$. 
\end{teo}
\textit{Proof:} Given an integral de Rham class $[\frac{1}{e}\omega]$ and a contractible open cover of $M$, $\mathcal{A}=\{A_j\}_{j\in I}$, on each open set $A_j$ the $2$-form $\omega$ is exact (Poincar\'{e} lemma), $\omega=\ud\vartheta_j$. The $1$-forms $\vartheta_j-\vartheta_k$ are closed over $A_j\cap A_k$; consequently, $\frac{1}{e}\vartheta_j-\frac{1}{e}\vartheta_k=\ud h_{jk}$, the mappings $A_j\cap A_k\ni p\mapsto\e^{ih_{jk}(p)}\in S^1$ give an element of $\check{H}^1(M;\mathcal{C}^\infty_{S^1})$ (since $[\frac{1}{e}\omega]$ is integral), and the potential $1$-forms $\frac{1}{e}\vartheta_j$ provide the hermitian connexion.\hfill $\blacksquare$\par\vspace{0.5em}

There is a converse of the previous statement. 

\begin{prop}\label{prop1}
 If the hermitian connexion $\nabla^{\omega}$ of a hermitian line bundle $(L,\escalar{\cdot}{\cdot})$ satisfies $\omega=-ie\cdot curv(\nabla^{\omega})$ for a real number $e\in\CR$, then the de Rham class $[\frac{1}{e}\omega]$ is integral.
\end{prop}
\textit{Proof:} Let $\mathcal{A}=\{A_j\}_{j\in I}$ be a contractible open cover of $M$ such that each $A_j$ is a trivialising neighbourhood of $L$ with unitary section $s_j$. On each $A_j$ the formula $\omega=e\ud\Theta_j$ holds: $\Theta_j$ is the potential $1$-form with respect to $s_j$ (lemma \ref{unitarypotential2}). As a consequence, over each intersection $\Theta_j-\Theta_k=\ud f_{jk}$, where $\e^{if_{jk}}\in C^\infty(A_j\cap A_k;S^1)$ are the transition functions (lemma \ref{unitarypotential}). Now, $\ud(f_{jk}+f_{kl}-f_{jl})=0$ implies that $f_{jk}+f_{kl}-f_{jl}=a_{jkl}\in\CR$ on $A_j\cap A_k\cap A_l$. The cocycle conditions of the hermitian line bundle, $\e^{if_{jk}}\e^{if_{kl}}=\e^{if_{jl}}$, imply that $\e^{ia_{jkl}}=1$; thus, $a_{jkl}\in 2\pi\CZ$ and $[\frac{1}{e}\omega]$ is integral.\hfill $\blacksquare$\par\vspace{0.5em}

As a consequence of theorem \ref{teo3} and proposition \ref{prop1}, flat hermitian line bundles are equivalent to hermitian line bundles admitting hermitian connexions whose curvatures are identically zero (flat connexions).

\begin{defi}The period between a closed $2$-form $\omega$ and a compact $2$-dimensional submanifold (embedded and without boundary) $N$ is given by $\mathrm{Per}([\omega],[N])=\integrall{N}{}{\omega}$.
\end{defi}

\begin{prop}
The period $\mathrm{Per}([\omega],[N])$ only depends on the de Rham cohomology class $[\omega]\in H_{dR}^2(M;\CR)$ and on the smooth singular homology class $[N]\in H^\infty_2(M;\CZ)$.
\end{prop}
\textit{Proof:} This is a straightforward application of Stoke's theorem. $N$ is a $2$-cycle, $\partial N=\emptyset$; ergo, if $\omega=\ud\vartheta$, 
\begin{equation*}
\integrall{N}{}{\omega}=\integrall{N}{}{\ud\vartheta}=\integrall{\partial N}{}{\vartheta}=0 \ ,
\end{equation*}and when $N$ is a $2$-boundary, $N=\partial \mathcal{N}$, 
\begin{equation*}
\integrall{N}{}{\omega}=\integrall{\partial \mathcal{N}}{}{\omega}=\integrall{\mathcal{N}}{}{\ud\omega}=0 \ .
\end{equation*}\hfill $\blacksquare$\par\vspace{0.5em} 

For an integral class $[\frac{1}{e}\omega]$, $\mathrm{Per}([\omega],[N])=e k([\omega],[N])$, where $k([\omega],[N])\in\CZ$.

\begin{teo}\label{teo5}If $[N_1],\dots,[N_{b_2(M)}]\in H^\infty_2(M;\CZ)$ generate the second smooth singular homology group of $M$, given a set of integers $\{k_1,\dots,k_{b_2(M)}\}\subset\CZ$, an integral class $[\frac{1}{e}\omega]\in H_{dR}^2(M;\CR)$ is uniquely defined by $\mathrm{Per}([\omega],[N_j])=k_je$.
\end{teo}
\textit{Proof:} This is a particular instance of de Rham's theorem.\hfill $\blacksquare$\par\vspace{0.5em}



\end{document}